\documentclass{aa}
\usepackage{graphicx}
\usepackage{amsmath}
\usepackage{booktabs}
\usepackage{microtype}
\usepackage[
colorlinks=true,
linkcolor=blue,
urlcolor=blue,
citecolor=blue,
anchorcolor=blue]
{hyperref}
\usepackage[varg]{txfonts}
\usepackage{natbib}
\bibpunct{(}{)}{;}{a}{}{,}

\newcommand{\st}[1]{_\text{#1}}
\newcommand{\FeH}{[\mathrm{Fe}/\mathrm{H}]}
\newcommand{\Teff}{T\st{eff}}
\newcommand{\K}{\rm\thinspace K}
\newcommand{\uHz}{\hbox{\rm\thinspace $\mu$Hz}}
\newcommand{\yr}{\rm\thinspace yr}
\newcommand{\Myr}{\rm\thinspace Myr}
\newcommand{\Gyr}{\rm\thinspace Gyr}
\newcommand{\Msun}{\hbox{$\rm\thinspace \text{M}_{\odot}$}}
\newcommand{\Rsun}{\hbox{$\rm\thinspace \text{R}_{\odot}$}}
\newcommand{\Lsun}{\hbox{$\rm\thinspace L_{\odot}$}}
\newcommand{\dex}{\rm\thinspace dex}
\newcommand{\eq}[1]{
\begin{equation}
#1
\end{equation}}

\begin{document}

\title{Surface-effect corrections for oscillation frequencies of
  evolved stars}%

\author{W.~H.~Ball\inst{1,2,3} 
  \and L.~Gizon\inst{2,1,4}}%
\institute{
  Institut f\"ur Astrophysik, Georg-August-Universit\"at G\"ottingen, 
  Friedrich-Hund-Platz 1, 37077 G\"ottingen, Germany
  \and
  Max-Planck-Institut f\"ur Sonnensystemforschung, 
  Justus-von-Liebig-Weg 3, 37077 G\"ottingen, Germany
  \and
  School of Physics and Astronomy, University of Birmingham, Edgbaston, Birmingham, B15 2TT, UK\\
  \email{wball@bison.ph.bham.ac.uk}
  \and
  Center for Space Science, NYUAD Institute, New York University Abu Dhabi, P.O.~Box 129188, Abu Dhabi, UAE}

\abstract{Accurate modelling of solar-like oscillators requires that
  modelled mode frequencies are corrected for the systematic shift
  caused by improper modelling of the near-surface layers, known as
  the surface effect.  Several parametrizations of the surface effect
  are now available but they have not yet been systematically compared
  with observations of stars showing modes with mixed g- and p-mode
  character.}
{We investigate how much additional uncertainty is introduced to
  stellar model parameters by our uncertainty about the functional
  form of the surface effect.  At the same time, we test whether any
  of the parametrizations is significantly better or worse at
  modelling observed subgiants and low-luminosity red giants.}%
{We model six stars observed by \emph{Kepler} that show clear mixed
  modes.  We fix the input physics of the stellar models and
  vary the choice of surface correction between five
  parametrizations.}%
{Models using a solar-calibrated power law correction consistently fit
  the observations more poorly than the other four corrections.
  Models with the remaining four corrections generally fit the
  observations about equally well, with the combined surface
  correction by Ball \& Gizon perhaps being marginally superior.  The
  fits broadly agree on the model parameters within about the
  $2\sigma$ uncertainties, with discrepancies between the modified
  Lorentzian and free power law corrections occasionally exceeding the
  $3\sigma$ level.  Relative to the best-fitting values, the total
  uncertainties on the masses, radii and ages of the stars are all
  less than $2$, $1$ and $6$ per cent, respectively.}
{A solar-calibrated power law, as formulated by Kjeldsen et al.,
  appears unsuitable for use with more evolved solar-like oscillators.
  Among the remaining surface corrections, the uncertainty in the
  model parameters introduced by the surface effects is about twice as
  large as the uncertainty in the individual fits for these six stars.
  Though the fits are thus somewhat less certain because of our
  uncertainty of how to manage the surface effect, these results also
  demonstrate that it is feasible to model the individual mode
  frequencies of subgiants and low-luminosity red giants, and hence
  also use these individual stars to help to constrain stellar
  models.}


\maketitle

\section{Introduction}

Space-based observations by CoRoT \citep{corot} and \textit{Kepler}
\citep{kepler} have led to the measurement and analysis of oscillation
mode frequencies in dozens of dwarf solar-like oscillators
\citep[e.g.][]{legacy1, legacy2}.  However, to fully exploit these
observations, one must correct for a systematic difference between
modelled and observed mode frequencies caused by improper modelling of
the near-surface layers of these stars.  This difference is known as
the ``surface effect'' or ``surface term'' and, if not corrected,
ultimately biases the stellar properties that are inferred.

Broadly speaking, the surface effect can be mitigated either by
applying some sort of correction to the computed frequencies or by
trying to model the stars and their oscillations better in the first
place.  Along the first line, \citet{kbcd2008} proposed to correct the
model frequencies using a solar-calibrated power law, which has
subsequently been used quite widely
\citep[e.g.][]{metcalfe2012,deheuvels2014}.  More recently,
\citet{ball2014} proposed a correction of the form
$\nu^3/\mathcal{I}$, possibly supplemented by a term of the form
$\nu^{-1}/\mathcal{I}$, where $\nu$ is the (cyclic) mode frequency and
$\mathcal{I}$ the mode inertia normalized at the photosphere.
Finally, \citet{sonoi2015} proposed a modified Lorentzian correction.

There has not been a systematic comparison of all of these
parametrized corrections in main-sequence stars.  \citet{ball2014}
compared their proposed corrections with the solar-calibrated power
law of \citet{kbcd2008} and found that their one-term correction
reproduced the solar surface effect better and also provided
better-fitting models for the G0V CoRoT target HD~52265.  The
additional term did not improve the fit significantly.  They showed
that the solar-calibrated power-law correction tends to overpredict
the magnitude of the surface correction at frequencies below about
$2500\uHz$ or above $3500\uHz$ and a similar trend was the main reason
for the poor fits to HD~52265.

\citet{schmitt2015} compared the surface corrections of
\citet{kbcd2008} and \citet{ball2014}, as well as a scaled solar
surface term (with or without an additive constant), by fitting
frequency differences between different published solar models,
between solar models calibrated with different model atmospheres, and
between stellar models in a large grid before and after modifying
their near-surface structure.  The two-term correction by
\citet{ball2014} was best at fitting the differences between the
published solar models and only marginally worse than a scaled solar
surface term, with an additive constant, at fitting the differences
between solar models calibrated with different atmospheres.  For the
grid of stellar models, \citet{schmitt2015} found that, depending on
how they modified the near-surface layers, either all the surface
corrections performed similarly well or the two-term correction by
\citet{ball2014} was clearly superior.  The modified Lorentzian of
\citet{sonoi2015} was not yet published and thus not considered in
their comparison.

Along a similar line to parametrizing the surface correction,
\citet{roxburgh2003} proposed to mitigate the surface effect by
instead considering ratios of frequency differences, specifically
constructed to cancel out the individual modes' sensitivities to the
near-surface layers.  \citet{oti2005} demonstrated that these
``separation ratios'' are mostly sensitive to a star's core and they
have also seen frequent use \citep[e.g.][]{reese2016,legacy2}.
\citet{roxburgh2015,roxburgh2016} has suggested new methods based on
the same underlying principles but these are too new to have seen use.

The last few years have seen a renewed effort towards the second
approach: improving the models of the near-surface layers, usually by
supplementing the stellar models with averaged structures from
three-dimensional radiation hydrodynamics (3D RHD) simulations.  The
process of replacing the near-surface layers of an existing model with
averaged 3D RHD simulation data is now usually referred to as
``patching'' and authors now present differences between ``patched''
and ``unpatched'' models.

\citet{rosenthal1999} computed frequency differences between
mixing-length models of the Sun's convective envelope before and after
patching.  They found the mode frequencies matched the observations
better (if one assumes that the perturbation to the turbulent pressure
behaves in the same way as the perturbation to the gas pressure) but a
significant---though smaller---systematic difference remained.
\citet{piau2014} presented the first frequency differences for patched
models of the whole Sun, corroborating the original conclusions of
\citet{rosenthal1999}.  Most recently, \citet{magic2016} compared the
frequencies with different magnetic field strengths included in the 3D
RHD simulations.  Finally, \citet{sonoi2015} and \citet{ball2016},
using different 3D RHD and stellar model codes, computed frequency
differences using patched models for stellar types other than the Sun,
broadly finding larger surface effects for hotter stars.
\citet{sonoi2015} also found smaller surface effects for less compact
stars (i.e. stars with lower surface gravities).

These calculations, however, are all limited to the structural part of
the surface effect.  It is also expected that some of the surface
effect is caused by the interaction between convection and pulsation.
\citet{houdek2016} considered these effects for a model of the solar
envelope, using a non-local theory of convection, and found that, once
full account is taken of various non-adiabatic and dynamical effects,
the difference between the modelled and observed frequencies improves
further still.  A residual difference persists but these calculations
can still be refined and bode well.

Most of the discussion of surface effects so far has considered the
Sun and dwarf stars without any mixed modes.  These are modes that
have oscillating components both in the outer layers of the star as
well as in the stellar core, separated by an evanescent region.  The
outer oscillating components correspond to p-modes, as observed in
dwarf stars like the Sun, whereas the inner oscillating components
correspond to g-modes.  In unevolved stars, like the Sun, the p-mode
and g-mode oscillations are confined to distinct ranges of frequencies
but, as a star evolves, the p-mode frequencies decrease while the
g-mode frequencies increase, and it becomes possible for the modes to
couple into a mixed mode with minimal damping between.  Such a mode
then has an observable amplitude at the surface and its mixed nature
is revealed by the mode frequency's deviation from the usual pattern
for p-modes.

Evolved red giants show dense spectra of mixed modes that are being
exploited to infer a tremendous amount of information about the stars'
cores \citep[see e.g.][for a recent review]{hekker2016} but smaller
numbers of mixed modes are also detected in subgiants and
low-luminosity red giants.  These modes are potentially problematic
for pipelines that are now standard for modelling dwarf solar-like
oscillators, to the extent that stars in these samples that show too
many mixed modes are omitted \citep[e.g.][]{metcalfe2014,legacy1}.
Modelling individual mixed modes in more evolved stars is even more
challenging.  Recent efforts typically use derived parameters
  (e.g. the asymptotic period spacing) instead of the individual mixed
  mode frequencies \citep[e.g.][]{perez2016}.

Our current interest is the nature of the surface effect in these
stars and their mixed modes.  The mixed modes' frequencies are
determined in part by the stellar core, which is presumably oblivious
to the star's near-surface structure.  We therefore expect that these
modes will show smaller frequency shifts as a result of the surface
effects but this has not yet been tested, and it is not obvious that
the parametrizations that have worked for dwarfs will continue to work
in more evolved stars.

Here, we consider surface corrections in six subgiant stars,
originally studied by \citet{deheuvels2014} precisely for their mixed
modes: KIC~12508433, KIC~8702606, KIC~5689820, KIC~8751420,
KIC~7799349 and KIC~9574283.  We compare best-fitting stellar models
with five different surface corrections with the aim of determining
how much uncertainty is induced on the stellar parameters by our
uncertainty about the surface effect, and to see if one correction
might be obviously better (or worse) than the others.  We do not
consider the correction methods proposed by \citet{roxburgh2003} and
\citet{roxburgh2015,roxburgh2016} because the underlying assumption of
the oscillations being in one cavity is not satisfied.

\section{Methods}
\label{s:methods}

\subsection{Stellar models}
\label{s:models}

We computed stellar models using the Modules for Experiments in
Stellar Astrophysics
\citep[MESA\footnote{\url{http://mesa.sourceforge.net}}, revision
7624;][]{paxton2011, paxton2013, paxton2015}.  Opacities are taken at
high- and low-temperatures from the tables of the OPAL collaboration
\citep{iglesias1996} and \citet{ferguson2005}, respectively.  The
equation of state is MESA's default, the relevant part of which is
principally based on the OPAL EOS \citep{rogers2002}.  Nuclear
reaction rates are taken from the NACRE tables \citep{angulo1999} or,
if not available there, from the tables by \citet{caughlan1988}.  For
the specific reactions ${}^{14}\mathrm{N}(p,\gamma){}^{15}\mathrm{O}$
and ${}^{12}\mathrm{C}(\alpha,\gamma){}^{16}\mathrm{O}$, we use
revised rate by \citet{imbriani2005} and \citet{kunz2002} (though the
latter is not relevant here).  Convection is described by
mixing-length theory \citep{boehm-vitense1958} as presented in
\citet{cox1968}.  Gravitational settling is implemented according to
\citet{thoul1994}; radiative levitation is neglected.  The solar metal
mixture is that of \citet{grevesse1998}.  Finally, the surface
boundary condition is determined by integrating a standard
Eddington-grey atmosphere from an optical depth of $\tau=10^{-4}$ to
the photospheric value of $\tau=2/3$ and the atmospheric structure is
included in the stellar model when computing the mode frequencies.
We computed linear adiabatic mode frequencies using the Aarhus
adiabatic pulsation code \citep[ADIPLS,][]{adipls}.  All models were
remeshed to contain 4800 meshpoints before the frequencies were
computed (from typically about 2000 meshpoints during the evolution).
The remeshing used the remeshing routine bundled with ADIPLS
\citep{jcd1991}, which uses the asymptotic behaviour of the modes to
redistribute points more evenly over the displacement eigenfunction.
Without remeshing, oscillation codes are known to sometimes miss mixed
modes.

\subsection{Surface terms}
\label{s:corrs}

We modelled the stars using five different surface corrections, which
each provide corrected model frequencies $\nu\st{cor}$ in terms of the
uncorrected model frequencies $\nu\st{mdl}$ such that they should
better match the observed frequencies $\nu\st{obs}$.

First, we used the solar-calibrated power law of \citet{kbcd2008},
with the power-law index fixed at a value of $p_1=5.00$ using a solar
calibration with the same input physics as above.  This is a typical
value for the parameter: \citet{kbcd2008} themselves found $p_1=4.90$.
The correction is then \eq{r\nu\st{cor}-\nu\st{mdl}=\frac{p_0}{Q}
  \left(\frac{\nu\st{mdl}}{\nu\st{ref}}\right)^{p_1}\label{e:kbcd}}%
where $r$ is the ratio of the square roots of the modelled and
observed mean densities and $\nu\st{ref}$ is a reference frequency,
taken here to be $\nu\st{max}$, the frequency of maximum oscillation
power, determined from the scaling relation \citep{kjeldsen1995}
\eq{\nu\st{max}=\frac{g}{g_\odot}
  \left(\frac{T\st{eff}}{T_{\mathrm{eff},\odot}}\right)^{-1/2}\nu_{\mathrm{max},\odot}}%
with $\log g_\odot=4.438$, $T_{\mathrm{eff},\odot}=5777\K$ and
$\nu_{\mathrm{max},\odot}=3090\uHz$.  The coefficient $p_0$ is
determined in essence by linear regression \citep[see][for
details]{kbcd2008}.  In eq.~(\ref{e:kbcd}), $Q$ is the ratio between the
mode inertia of the mode divided by the inertia that a radial mode
would have at the same frequency, determined by linear interpolation.
This factor has long been used to capture the variation of the surface
effects with mode inertia \citep{jcd1986}.

Second, we used the corrections proposed by \citet{ball2014}: their
``cubic'' correction
\eq{\nu\st{cor}-\nu\st{mdl}=a_3\left(\frac{\nu\st{mdl}}{\nu\st{ac}}\right)^3/\mathcal{I}\label{e:cube}}%
and their ``combined'' correction
\eq{\nu\st{cor}-\nu\st{mdl}=\left[a_{-1}\left(\frac{\nu\st{mdl}}{\nu\st{ac}}\right)^{-1}+a_3\left(\frac{\nu\st{mdl}}{\nu\st{ac}}\right)^3\right]/\mathcal{I}\label{e:both}}%
Here, $\mathcal{I}$ is the mode inertia normalized at the photosphere,
$\nu\st{ac}$ is the acoustic cut-off frequency, scaled from a solar
value of $5000\uHz$ \citep{jimenez2011} using the scaling relations,
and the parameters $a_{-1}$ and $a_3$ are determined by linear
regression to minimize the differences between all the observed and
modelled mode frequencies.  The acoustic cut-off is used purely to
rescale the coefficients $a_{-1}$ and $a_3$.  The correction is based
on the asymptotic behaviour of the eigenfunctions near the surface.
From the variational principle for linear adiabatic oscillations,
\citet{gough1990} showed that a simple sound speed perturbation would
cause a frequency shift of the form $\nu^3/\mathcal{I}$, whereas a
perturbation to the pressure scale height would cause a shift of the
form $\nu^{-1}/\mathcal{I}$.  \citet{ball2014} showed that functions
of this form give a better fit to the known solar surface effect than
a power law.

Third, we considered a modified Lorentzian correction
\eq{\nu\st{cor}-\nu\st{mdl}=\frac{s_0\nu\st{max}}{Q}
  \left[1-\frac{1}{1+(\nu\st{mdl}/\nu\st{max})^{s_1}}\right]\label{e:sonoi}}%
as suggested by \citet{sonoi2015}.  Rather than use their calibrations
of $s_0$ and $s_1$ to frequency changes between their patched and
unpatched models, we treated them as free parameters, optimized for
each stellar model by the Newton--Rhapson method to fit the observed
frequencies using all of the observed modes.  \citet{sonoi2015} only
considered radial modes in formulating eq.~(\ref{e:sonoi}) so we have
chosen to include the factor $Q$, as in the correction by
\citet{kbcd2008}.  Also, after performing a large number of fits, we
found that the best-fitting models for stars B and D had positive
values of $s_0$ and negative values of $s_1$, which corresponds to a
surface correction of opposite sign to the Sun and decreasing in
magnitude with increasing frequency.  We regard this as unphysical and
subsequently restricted the fits to have $s_1>0$.

Finally, we also included a free power law, as in eq.~(\ref{e:kbcd})
but with both $p_0$ and $p_1$ as free parameters and with $r$ fixed to
$1$.  The quality of these fits gives us some idea of how much we are
gaining from more complicated models of the surface term.  As with the
modified Lorentzian, solutions with $p_1<0$ correspond to surface
corrections that decrease in magnitude with increasing frequency, so
we restricted our best-fit models to those with $p_1>0$.

\begin{table*}
\centering
\caption{Global seismic and non-seismic parameters for the six
  subgiants studied in this article.  The effective temperatures and
  metallicities are taken from the spectroscopic values in
  Table~10 of \citet{deheuvels2014}.  The large separations
  $\Delta\nu$ and frequencies of maximum oscillation power
  $\nu\st{max}$ are from their Table~1.}
\begin{tabular}{cr@{\,}rr@{\,$\pm$\,}lr@{\,$\pm$\,}lr@{\,$\pm$\,}lr@{\,$\pm$\,}l}
\toprule
Star & \multicolumn{2}{c}{KIC} & 
\multicolumn{2}{c}{$\Teff/\K$} & \multicolumn{2}{c}{$\FeH$} & 
\multicolumn{2}{c}{$\Delta\nu/\uHz$} & \multicolumn{2}{c}{$\nu\st{max}/\uHz$} \\
\midrule
A & KIC & 12508433 & $5248$ & $130$ & $ 0.25$ & $0.23$ & $45.3$ & $0.2$ & $793$ & $21$ \\
B & KIC &  8702606 & $5540$ & $60 $ & $-0.09$ & $0.06$ & $39.9$ & $0.4$ & $664$ & $14$ \\
C & KIC &  5689820 & $4978$ & $167$ & $ 0.24$ & $0.16$ & $41.0$ & $0.3$ & $695$ & $15$ \\
D & KIC &  8751420 & $5264$ & $60 $ & $-0.15$ & $0.06$ & $34.7$ & $0.4$ & $598$ & $14$ \\
E & KIC &  7799349 & $5115$ & $60 $ & $ 0.41$ & $0.06$ & $33.7$ & $0.4$ & $561$ & $ 8$ \\
F & KIC &  9574283 & $5120$ & $55 $ & $-0.40$ & $0.08$ & $30.0$ & $0.5$ & $455$ & $ 8$ \\
\bottomrule
\end{tabular}
\label{t:data}
\end{table*}

\begin{figure}
\centering
\includegraphics[width=85mm]{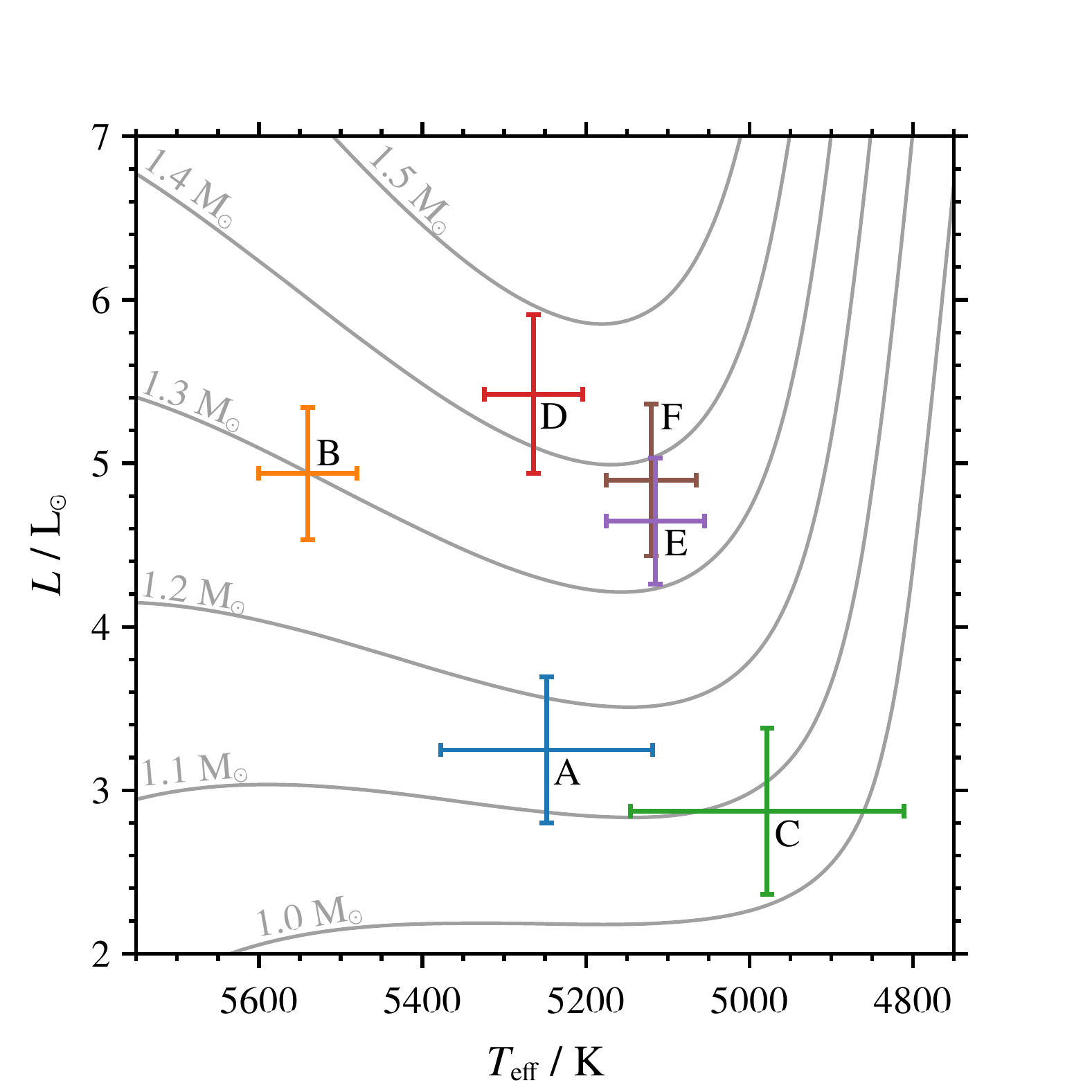}
\caption{The locations of the six stars under study in the
  Hertzsprung--Russell diagram using luminosities that have
  been computed using scaling relations.  The solid lines are
  evolutionary tracks for stellar models using a solar-calibrated
  mixing-length parameter and composition for mass from $1.0$ to
  $1.5\Msun$ in steps of $0.1\Msun$.  The sample of stars are all
  located towards the end of the subgiant branch or at the base of the
  red giant branch.}
\label{f:HR}
\end{figure}

\begin{table}
\centering
\caption{Mode frequencies that were excluded in the analysis because
  of poor initial fits.}
\begin{tabular}{ccc}
\toprule
Star & Frequency$/\uHz$ & $\ell$ \\
\midrule
B & $491.070\pm0.024$ & $0$ \\
C & $511.857\pm0.011$ & $2$ \\
E & $511.478\pm0.032$ & $2$ \\
F & $398.982\pm0.021$ & $1$ \\
\bottomrule
\end{tabular}
\label{t:excluded}
\end{table}

\subsection{Sample of stars}

We modelled the six stars studied by \citet{deheuvels2014}, who
selected these subgiants and low-luminosity red giants to invert for
their rotational profiles.  They are KIC~12508433, KIC~8702606,
KIC~5689820, KIC~8751420, KIC~7799349 and KIC~9574283, labelled
hereafter by letters A through F, in decreasing order of the surface
gravity inferred from the global seismic parameters \citep[as
in][]{deheuvels2014}.  Basic data is given in Table~\ref{t:data} and
their positions in the Hertzsprung--Russell diagram are shown in
Fig.~\ref{f:HR}, where the luminosities have been inferred from
scaling relations.  

Rotational inversions require several features of the observations.
The stars have high signal-to-noise ratios and clear mixed modes, for
which the rotational splitting is well-resolved.  The stars could
feasibly be modelled to identify which frequency corresponds to which
mode.  As stars ascend the red giant branch, they develop an ever
denser spectrum of mixed modes and matching the correct modelled mode
to the observed mode frequency becomes ambiguous, which ruled out more
evolved stars.  Our study also requires stars with clear mixed modes
and high signal-to-noise ratio in all the observed modes, so it is
clear that these six stars are suitable.

At first, we used all of the frequencies and their uncertainties given
by \citet{deheuvels2014}.  We found that we could not reasonably model
stars B, C, E and F as long as we included the worst-fitting
modes, which contributed more than half of the total $\chi^2$ for each
star, and sometimes nearly all of it.  We therefore removed these
modes from our data.  The frequencies and angular degrees of the
excluded modes are given in Table~\ref{t:excluded}.  Of course,
rejecting the data that fits worst will inevitably give better fits.
However, in these cases we specifically found that the poorly fitting
modes dominated $\chi^2$ so much that they were clearly biasing the
results away from models that were much better able to reproduce the
other mode frequencies.  We kept other mode frequencies that appear
consistently discrepant, since those modes did not appear to bias our
results away from models that could fit the remaining data. e.g. the
highest-frequency mode in star F ($\nu=581.219\uHz$).

\subsection{Fitting method}
\label{s:fits}

We fit the stellar models using essentially the same method presented
in other studies to which we contributed best fit models
\citep[e.g.][]{appourchaux2015, reese2016}.  We optimized the total
$\chi^2$ of the observations, defined by
\eq{\chi^2=\sum_{i=1}^{N\st{obs}}
  \left(\frac{y_{\mathrm{obs},i}-y_{\mathrm{mdl},i}}{\sigma_i}\right)^2\label{e:chi2}}%
where $y_{\mathrm{obs},i}$, $y_{\mathrm{mdl},i}$ and $\sigma_i$ are
the observed value, modelled value and observed uncertainty of the
$i$-th observable.  Here, these are the effective temperature $\Teff$,
the metallicity $\FeH\st{s}$ (see Table~\ref{t:data}) and the
individual mode frequencies.  Specifically, we did not weight any part
of $\chi^2$ by any additional factor.

We first estimated some stellar parameters using coarse grid-based
modelling before moving onto an iterative method.  For each set of
mass $M$, initial helium abundance $Y_0$, initial metallicity $\FeH_0$
and mixing-length parameter $\alpha$, we started an evolutionary track
from a chemically-homogeneous pre-main-sequence model with central
temperature $9\times10^5\K$.  The timestep was gradually reduced as
the stellar model first matched the spectroscopic parameters and then
the radial mode frequencies.  At this point, all the mode frequencies
were computed and the total $\chi^2$ evaluated.  The main difference
between the current and previous optimizations (all for main-sequence
or near-main-sequence stars) was that the minimum timestep was reduced
to as little as $3\,200\yr$ in order to compute models during the
rapid evolution of the mixed mode frequencies.

Next, we performed iterations of the Nelder--Mead downhill simplex
method \citep{nelder1965}, with additional linear extrapolations to
explore parameter space, and recorded each set of model parameters
that were tried.  When the next step of the downhill simplex would
have been a contraction step, we tried to generate better-fitting
model parameters by various methods, including: linear extrapolations
from random subsamples of the sample so far; small, dense grids
spanning the present estimate of the $1\sigma$ to $5\sigma$ confidence
regions; or random uniform samples within the $1\sigma$ to $5\sigma$
confidence regions.  When nothing seemed to improve the fit any
further, the best-fitting model parameters were used for an
uninterrupted standard downhill simplex to check that we had found at
least a locally-optimal model.

The above process, though somewhat haphazard, was aimed at preventing
convergence on a local minimum, which we sometimes find is a problem.
Though we cannot guarantee that our best-fitting models are not local
minima, our extensive searches around the best-fit parameters give us
some confidence that they are probably not local minima.  We
determined uncertainties from ellipsoids bounding surfaces of constant
$\chi^2$, finding ellipsoids that would simultaneously enclose all
parameters in the sample within the region appropriate to the
corresponding value of $\chi^2$.  In other words, given a best fit
with $\chi^2=\chi^2_0$, we required that if a sample with $\chi^2_0+1$
was contained in the $1\sigma$ ellipsoid, a sample with $\chi^2_0+4$
had to simultaneously be contained in the corresponding $2\sigma$
ellipsoid.  To determine the uncertainties of derived parameters
(e.g. radius or effective temperature) we performed a linear fit of
the derived parameters relative to the model parameters and used
linear propagation of uncertainties.

\begin{figure*}
\centering
\includegraphics[width=170mm]{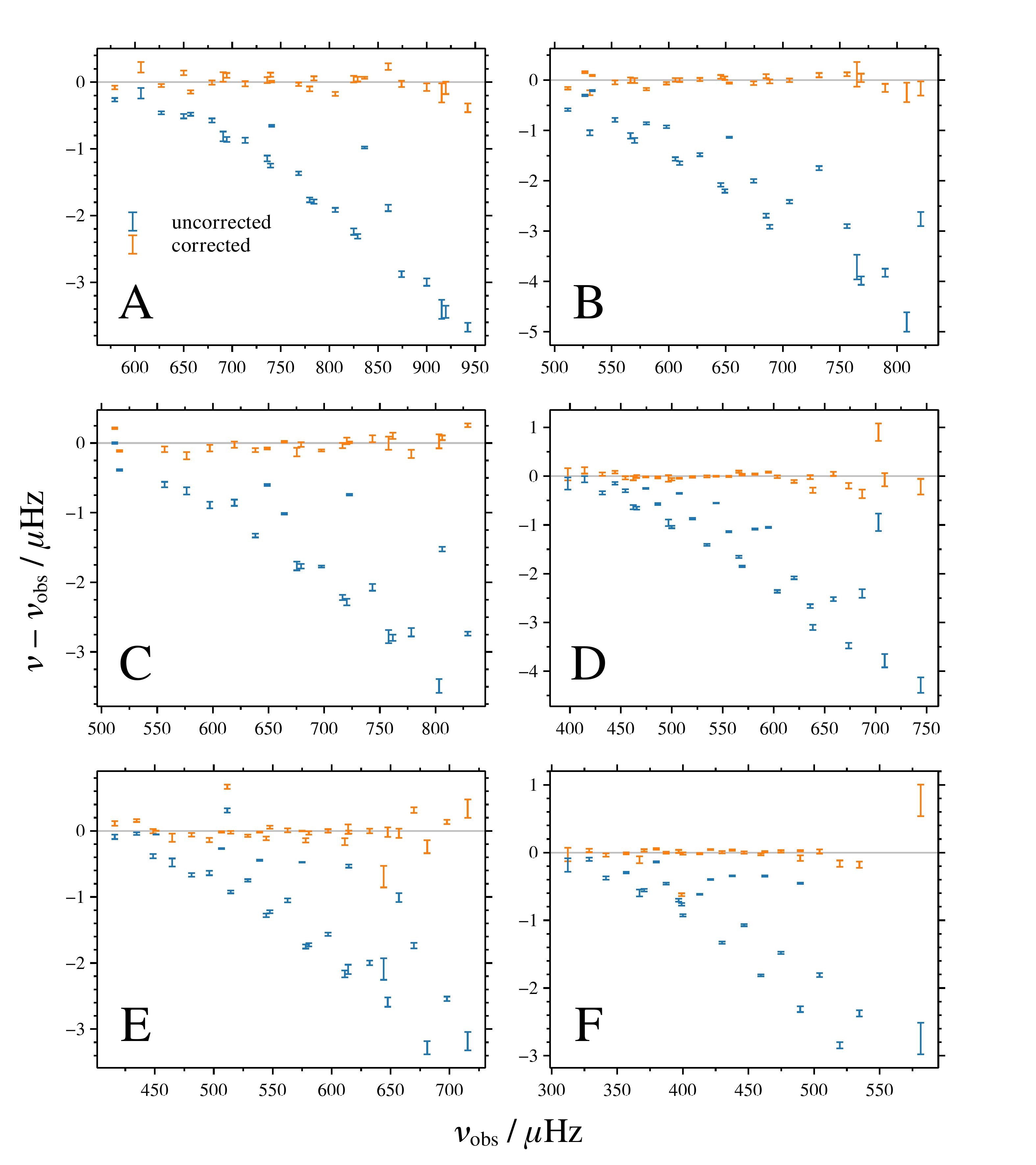}
\caption{Differences between modelled and observed frequencies plotted
  against observed frequency, for best-fitting models using the cubic
  correction of \citet{ball2014}.  Each panel corresponds to one of
  the stars A to F.  The blue points are frequency differences before
  the correction is applied; the orange points after.  The error bars
  are the observed uncertainties in both cases.}
\label{f:cube}
\end{figure*}

\begin{figure*}
\centering
\includegraphics[width=170mm]{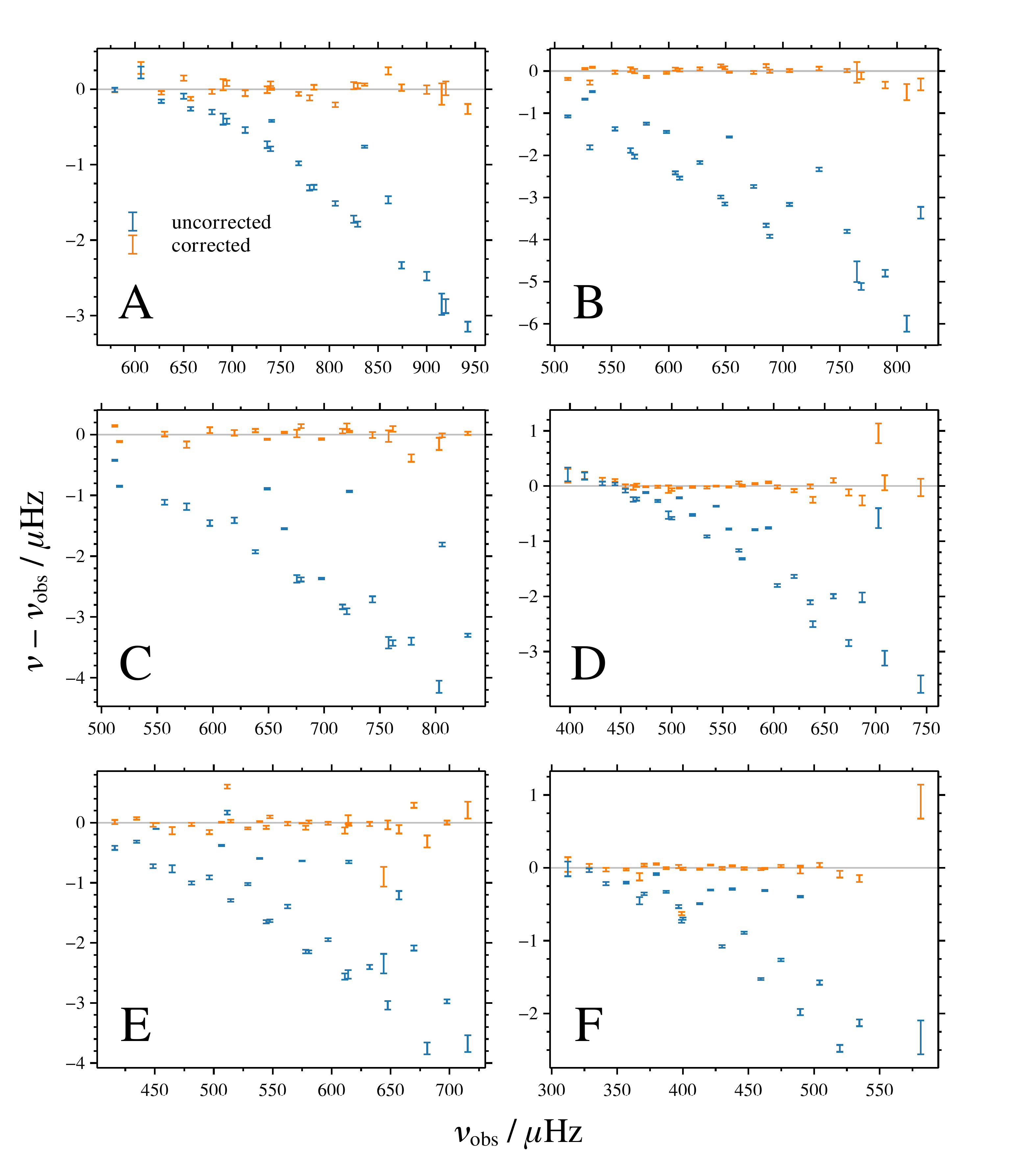}
\caption{As in Fig.~\ref{f:cube} but for best-fitting models using the
  combined correction of \citet{ball2014}.}
\label{f:both}
\end{figure*}

\begin{figure*}
\centering
\includegraphics[width=170mm]{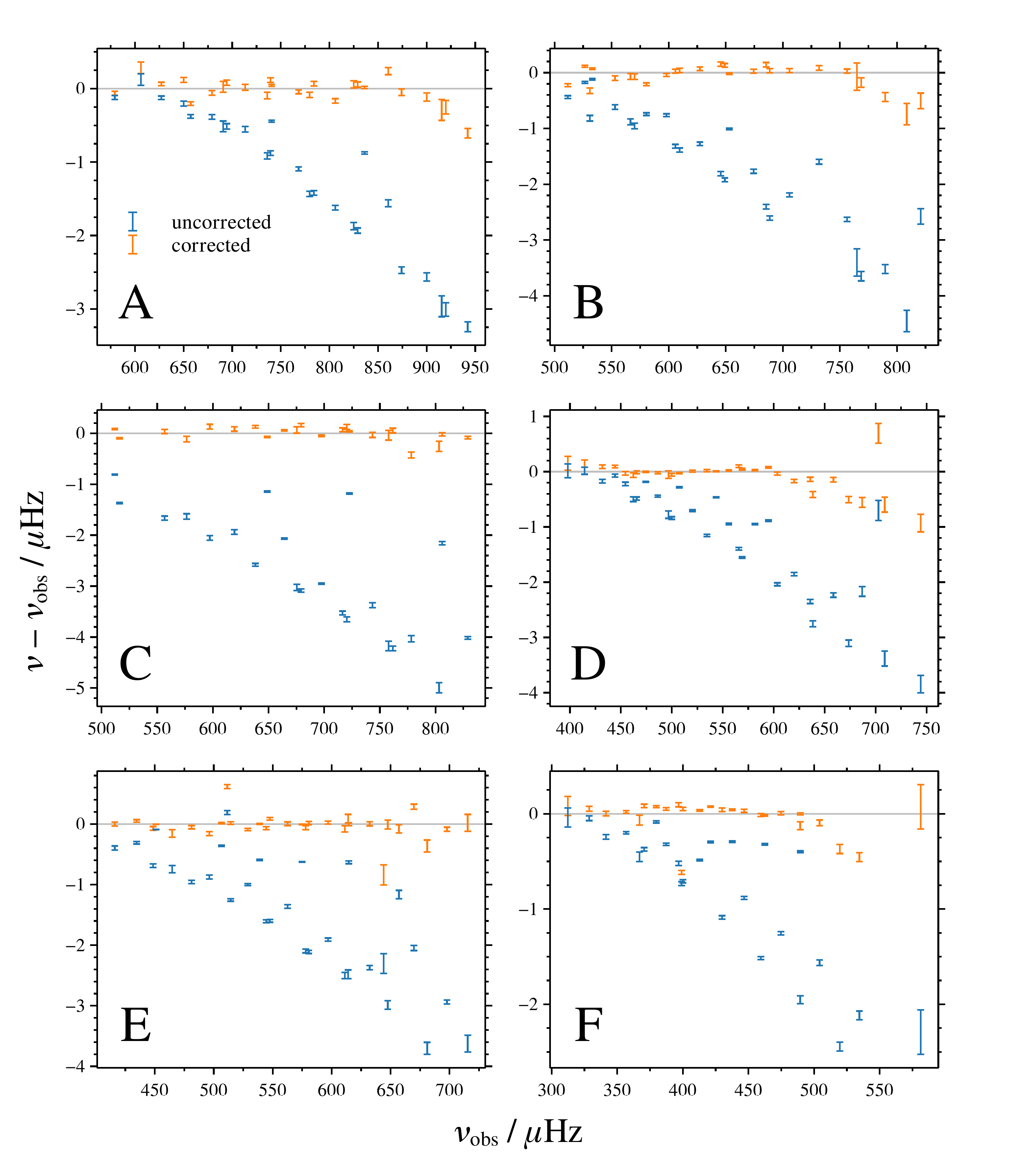}
\caption{As in Fig.~\ref{f:cube} but for best-fitting models using the
  modified Lorentzian correction proposed by \citet{sonoi2015}.}
\label{f:sonoi}
\end{figure*}

\begin{figure*}
\centering
\includegraphics[width=170mm]{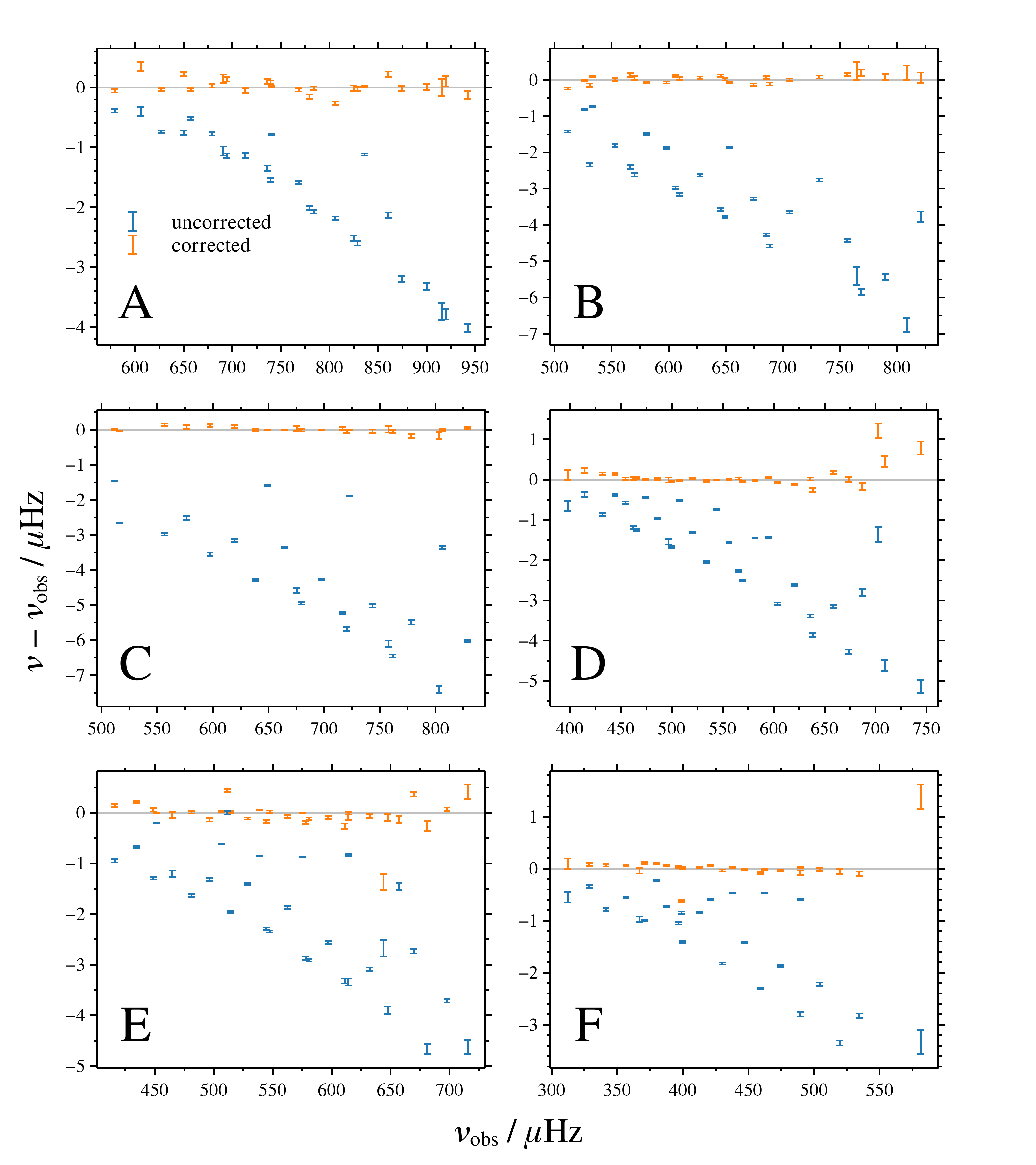}
\caption{As in Fig.~\ref{f:cube} but for best-fitting models using a free
  power law.}
\label{f:plaw}
\end{figure*}

\begin{figure*}
\centering
\includegraphics[width=170mm]{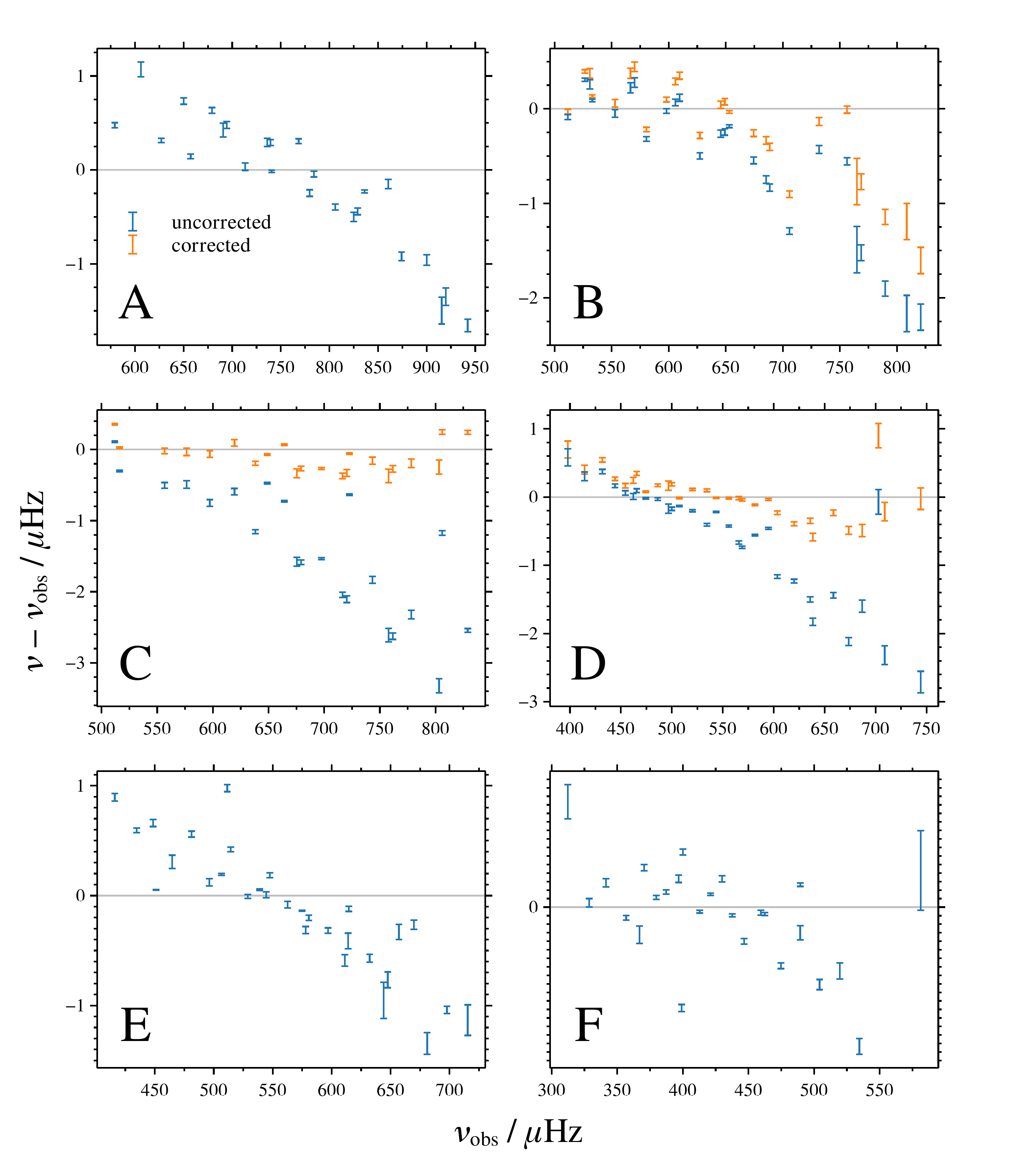}
\caption{As in Fig.~\ref{f:cube} but for best-fitting models using the
  solar-calibrated power-law correction of \citet{kbcd2008}.}
\label{f:kbcd}
\end{figure*}

\section{Results for individual stars}

\begin{table*}
\centering
\caption{Model parameters with uncertainties for fits for each of the six stars
  and five surface corrections.  The surface corrections `c', `b', `s',
  `p' and `k' refer to the cubic and combined terms of \citet{ball2014},
  the modified Lorentzian of \citet{sonoi2015}, a free power law and the
  solar-calibrated power law by \citet{kbcd2008}.  The surface correction 
  parameters in each of the five cases are $a_3/10^{-7}\uHz$ (eq.~\ref{e:cube}), 
  $a_{-1}/10^{-8}\uHz$ and $a_3/10^{-7}\uHz$ (eq.~\ref{e:both}),
  $s_0/10^{-3}\uHz$ and $s_1$ (eq.~\ref{e:sonoi}), $p_0/\uHz$ and $p_1$ (eq.~\ref{e:kbcd}, $r\equiv1$), 
  and finally $p_0/r/\uHz$ (eq.~\ref{e:kbcd}).  The model parameters are the age $t$, mass $M$, initial helium
  abundance $Y_0$, initial metallicity $\FeH_0$ and mixing-length parameter $\alpha$.}
\begin{tabular}{cccccrcrr}
\toprule
Name & Corr. & $t/\Gyr$ & $M/\Msun$ & $Y_i$ & \multicolumn{1}{c}{$\FeH_i$} & $\alpha$ & \multicolumn{2}{c}{Surface correction parameters} \\
\midrule
 & c & $  5.15 \pm   0.13$ & $  1.290 \pm   0.012$ & $  0.256 \pm   0.007$ & $  0.093 \pm   0.033$ & $  2.07 \pm   0.05$ & \multicolumn{2}{c}{$ -4.70 \pm   0.13$} \\
 & b & $  5.07 \pm   0.13$ & $  1.293 \pm   0.010$ & $  0.255 \pm   0.005$ & $  0.081 \pm   0.029$ & $  2.09 \pm   0.05$ & $  2.00 \pm   0.61$ & $ -4.75 \pm   0.07$ \\
A & s & $  5.59 \pm   0.14$ & $  1.262 \pm   0.009$ & $  0.271 \pm   0.005$ & $  0.191 \pm   0.027$ & $  1.92 \pm   0.04$ & $ -4.18 \pm   0.18$ & $ 10.70 \pm   0.68$ \\
 & p & $  4.82 \pm   0.09$ & $  1.310 \pm   0.006$ & $  0.244 \pm   0.003$ & $  0.015 \pm   0.021$ & $  2.20 \pm   0.04$ & $ -2.08 \pm   0.15$ & $  3.96 \pm   0.22$ \\
 & k & $  4.98 \pm   0.03$ & $  1.438 \pm   0.005$ & $  0.178 \pm   0.003$ & $ -0.067 \pm   0.007$ & $  2.22 \pm   0.01$ & \multicolumn{2}{c}{$ -0.00 \pm   0.02$} \\
\\
 & c & $  3.86 \pm   0.02$ & $  1.260 \pm   0.010$ & $  0.266 \pm   0.006$ & $ -0.184 \pm   0.016$ & $  2.06 \pm   0.03$ & \multicolumn{2}{c}{$ -6.98 \pm   0.21$} \\
 & b & $  3.84 \pm   0.03$ & $  1.248 \pm   0.009$ & $  0.270 \pm   0.004$ & $ -0.199 \pm   0.017$ & $  2.10 \pm   0.03$ & $ -3.95 \pm   0.73$ & $ -7.14 \pm   0.21$ \\
B & s & $  3.89 \pm   0.03$ & $  1.266 \pm   0.015$ & $  0.263 \pm   0.008$ & $ -0.179 \pm   0.020$ & $  2.04 \pm   0.03$ & $ -6.49 \pm   0.53$ & $  9.58 \pm   0.82$ \\
 & p & $  3.74 \pm   0.02$ & $  1.240 \pm   0.006$ & $  0.270 \pm   0.003$ & $ -0.248 \pm   0.015$ & $  2.23 \pm   0.03$ & $ -3.47 \pm   0.61$ & $  2.64 \pm   0.50$ \\
 & k & $  3.61 \pm   0.02$ & $  1.389 \pm   0.006$ & $  0.216 \pm   0.003$ & $ -0.246 \pm   0.006$ & $  2.22 \pm   0.01$ & \multicolumn{2}{c}{$ -0.41 \pm   0.04$} \\
\\
 & c & $  7.36 \pm   0.07$ & $  1.152 \pm   0.007$ & $  0.268 \pm   0.003$ & $  0.104 \pm   0.008$ & $  1.99 \pm   0.01$ & \multicolumn{2}{c}{$ -6.84 \pm   0.12$} \\
 & b & $  7.35 \pm   0.10$ & $  1.152 \pm   0.008$ & $  0.268 \pm   0.004$ & $  0.103 \pm   0.009$ & $  1.99 \pm   0.01$ & $ -0.27 \pm   0.62$ & $ -6.80 \pm   0.11$ \\
C & s & $  7.40 \pm   0.07$ & $  1.178 \pm   0.007$ & $  0.255 \pm   0.003$ & $  0.098 \pm   0.009$ & $  1.97 \pm   0.01$ & $ -5.91 \pm   0.32$ & $  9.05 \pm   0.41$ \\
 & p & $  6.79 \pm   0.10$ & $  1.116 \pm   0.010$ & $  0.285 \pm   0.004$ & $  0.046 \pm   0.014$ & $  2.12 \pm   0.02$ & $ -4.43 \pm   0.36$ & $  2.47 \pm   0.16$ \\
 & k & $  7.39 \pm   0.05$ & $  1.101 \pm   0.005$ & $  0.295 \pm   0.003$ & $  0.145 \pm   0.006$ & $  1.99 \pm   0.01$ & \multicolumn{2}{c}{$ -1.71 \pm   0.02$} \\
\\
 & c & $  4.25 \pm   0.05$ & $  1.286 \pm   0.010$ & $  0.226 \pm   0.006$ & $ -0.319 \pm   0.019$ & $  2.07 \pm   0.02$ & \multicolumn{2}{c}{$ -8.62 \pm   0.18$} \\
 & b & $  4.28 \pm   0.05$ & $  1.292 \pm   0.010$ & $  0.223 \pm   0.006$ & $ -0.310 \pm   0.017$ & $  2.04 \pm   0.02$ & $  3.22 \pm   0.93$ & $ -8.47 \pm   0.20$ \\
D & s & $  4.20 \pm   0.07$ & $  1.315 \pm   0.008$ & $  0.211 \pm   0.005$ & $ -0.357 \pm   0.014$ & $  2.10 \pm   0.02$ & $ -5.68 \pm   0.44$ & $  8.62 \pm   0.65$ \\
 & p & $  4.39 \pm   0.07$ & $  1.286 \pm   0.009$ & $  0.217 \pm   0.005$ & $ -0.352 \pm   0.012$ & $  2.09 \pm   0.02$ & $ -2.35 \pm   0.18$ & $  3.26 \pm   0.16$ \\
 & k & $  4.03 \pm   0.05$ & $  1.365 \pm   0.004$ & $  0.195 \pm   0.002$ & $ -0.371 \pm   0.007$ & $  2.11 \pm   0.01$ & \multicolumn{2}{c}{$ -0.76 \pm   0.01$} \\
\\
 & c & $  3.75 \pm   0.17$ & $  1.495 \pm   0.010$ & $  0.233 \pm   0.005$ & $  0.132 \pm   0.014$ & $  1.94 \pm   0.04$ & \multicolumn{2}{c}{$ -7.08 \pm   0.20$} \\
 & b & $  4.09 \pm   0.26$ & $  1.463 \pm   0.008$ & $  0.239 \pm   0.008$ & $  0.176 \pm   0.014$ & $  1.89 \pm   0.06$ & $ -2.92 \pm   0.92$ & $ -7.26 \pm   0.31$ \\
E & s & $  4.05 \pm   0.13$ & $  1.463 \pm   0.009$ & $  0.240 \pm   0.004$ & $  0.177 \pm   0.011$ & $  1.89 \pm   0.03$ & $ -7.31 \pm   0.28$ & $  7.28 \pm   0.39$ \\
 & p & $  4.12 \pm   0.18$ & $  1.425 \pm   0.014$ & $  0.248 \pm   0.009$ & $  0.158 \pm   0.008$ & $  1.94 \pm   0.04$ & $ -2.79 \pm   0.73$ & $  2.77 \pm   0.66$ \\
 & k & $  3.99 \pm   0.04$ & $  1.639 \pm   0.004$ & $  0.183 \pm   0.001$ & $  0.182 \pm   0.005$ & $  1.73 \pm   0.01$ & \multicolumn{2}{c}{$  0.03 \pm   0.06$} \\
\\
 & c & $  7.63 \pm   0.20$ & $  1.040 \pm   0.012$ & $  0.245 \pm   0.009$ & $ -0.408 \pm   0.021$ & $  1.88 \pm   0.02$ & \multicolumn{2}{c}{$-13.44 \pm   0.73$} \\
 & b & $  7.65 \pm   0.20$ & $  1.033 \pm   0.012$ & $  0.249 \pm   0.007$ & $ -0.397 \pm   0.018$ & $  1.88 \pm   0.02$ & $  1.90 \pm   1.16$ & $-13.47 \pm   0.41$ \\
F & s & $  7.55 \pm   0.27$ & $  1.063 \pm   0.015$ & $  0.233 \pm   0.005$ & $ -0.423 \pm   0.015$ & $  1.87 \pm   0.02$ & $ -6.69 \pm   0.48$ & $  9.41 \pm   0.56$ \\
 & p & $  7.07 \pm   0.16$ & $  1.057 \pm   0.013$ & $  0.240 \pm   0.006$ & $ -0.472 \pm   0.012$ & $  1.96 \pm   0.02$ & $ -2.08 \pm   0.16$ & $  3.43 \pm   0.21$ \\
 & k & $  3.93 \pm   0.07$ & $  1.261 \pm   0.003$ & $  0.231 \pm   0.002$ & $ -0.478 \pm   0.005$ & $  2.35 \pm   0.01$ & \multicolumn{2}{c}{$  0.02 \pm   0.01$} \\
\\
\bottomrule
\end{tabular}

\label{t:mdl}
\end{table*}

\begin{table*}
\centering
\caption{Table of observed quantities and predictions for 
  each of the six stars with the five different surface corrections.
  The labelling is as in Table~\ref{t:mdl}, with an extra row `o' that
  gives the observed quantities.  The effective temperature $\Teff$ and
  surface metallicity $\FeH_s$ are as listed by \citet{deheuvels2014}.  
  The other observables---surface gravity $\log g$, luminosity $L$ and 
  radius $R$---are derived from the scaling relations.  The last two columns
  give the total $\chi^2$ (eq.~\ref{e:chi2}) and $\chi^2$ per degree of 
  freedom.}
\begin{tabular}{cclcrccrr}
\toprule
Name & Corr. & \multicolumn{1}{c}{$T\st{eff}/\K$} & $\log g$ & \multicolumn{1}{c}{$\FeH_s$} & $L/\Lsun$ & $R/\Rsun$ & $\chi^2$ & $\chi^2_r$ \\
\midrule
 & o & $5248 \pm  130$ & $  3.826 \pm   0.013$ & $  0.250 \pm   0.230$ & $  3.246 \pm   0.445$ & $  2.175 \pm   0.070$ & & \\
 & c& $5409 \pm  51$ & $  3.835 \pm   0.001$ & $  0.073 \pm   0.032$ & $  3.983 \pm   0.173$ & $  2.275 \pm   0.008$ & $229.5$ & $ 10.9$\\
 & b& $5434 \pm  47$ & $  3.834 \pm   0.001$ & $  0.061 \pm   0.028$ & $  4.065 \pm   0.158$ & $  2.277 \pm   0.006$ & $204.5$ & $ 10.2$\\
A & s& $5255 \pm  45$ & $  3.831 \pm   0.001$ & $  0.170 \pm   0.026$ & $  3.495 \pm   0.141$ & $  2.258 \pm   0.006$ & $325.6$ & $ 16.3$\\
 & p& $5545 \pm  37$ & $  3.837 \pm   0.001$ & $ -0.005 \pm   0.021$ & $  4.443 \pm   0.126$ & $  2.287 \pm   0.004$ & $237.9$ & $ 11.9$\\
 & k& $5532 \pm  10$ & $  3.848 \pm   0.000$ & $ -0.086 \pm   0.007$ & $  4.706 \pm   0.035$ & $  2.364 \pm   0.003$ & $4467.4$ & $212.7$\\
\\
 & o & $5540 \pm   60$ & $  3.761 \pm   0.010$ & $ -0.090 \pm   0.060$ & $  4.935 \pm   0.404$ & $  2.413 \pm   0.075$ & & \\
 & c& $5708 \pm  21$ & $  3.757 \pm   0.001$ & $ -0.207 \pm   0.016$ & $  5.764 \pm   0.103$ & $  2.458 \pm   0.006$ & $308.6$ & $ 14.0$\\
 & b& $5746 \pm  26$ & $  3.757 \pm   0.001$ & $ -0.223 \pm   0.017$ & $  5.865 \pm   0.112$ & $  2.447 \pm   0.006$ & $258.3$ & $ 12.3$\\
B & s& $5690 \pm  28$ & $  3.758 \pm   0.002$ & $ -0.201 \pm   0.020$ & $  5.713 \pm   0.130$ & $  2.462 \pm   0.010$ & $401.1$ & $ 19.1$\\
 & p& $5846 \pm  25$ & $  3.756 \pm   0.001$ & $ -0.272 \pm   0.015$ & $  6.250 \pm   0.101$ & $  2.440 \pm   0.005$ & $297.2$ & $ 14.2$\\
 & k& $5811 \pm   9$ & $  3.769 \pm   0.001$ & $ -0.268 \pm   0.006$ & $  6.636 \pm   0.044$ & $  2.544 \pm   0.004$ & $2510.7$ & $114.1$\\
\\
 & o & $4978 \pm  167$ & $  3.758 \pm   0.013$ & $  0.240 \pm   0.160$ & $  2.870 \pm   0.511$ & $  2.268 \pm   0.074$ & & \\
 & c& $5098 \pm  12$ & $  3.773 \pm   0.001$ & $  0.084 \pm   0.008$ & $  3.231 \pm   0.029$ & $  2.307 \pm   0.005$ & $ 19.7$ & $  1.2$\\
 & b& $5100 \pm  14$ & $  3.773 \pm   0.001$ & $  0.084 \pm   0.009$ & $  3.236 \pm   0.040$ & $  2.307 \pm   0.006$ & $ 19.2$ & $  1.2$\\
C & s& $5075 \pm  13$ & $  3.776 \pm   0.001$ & $  0.079 \pm   0.009$ & $  3.223 \pm   0.033$ & $  2.325 \pm   0.005$ & $ 36.8$ & $  2.3$\\
 & p& $5247 \pm  27$ & $  3.770 \pm   0.001$ & $  0.027 \pm   0.014$ & $  3.535 \pm   0.058$ & $  2.278 \pm   0.008$ & $ 53.8$ & $  3.4$\\
 & k& $5114 \pm  10$ & $  3.767 \pm   0.001$ & $  0.125 \pm   0.006$ & $  3.174 \pm   0.021$ & $  2.272 \pm   0.004$ & $457.1$ & $ 26.9$\\
\\
 & o & $5264 \pm   60$ & $  3.704 \pm   0.011$ & $ -0.150 \pm   0.060$ & $  5.426 \pm   0.486$ & $  2.802 \pm   0.098$ & & \\
 & c& $5433 \pm  17$ & $  3.688 \pm   0.001$ & $ -0.333 \pm   0.019$ & $  5.667 \pm   0.095$ & $  2.690 \pm   0.007$ & $224.7$ & $  8.6$\\
 & b& $5407 \pm  19$ & $  3.688 \pm   0.001$ & $ -0.324 \pm   0.017$ & $  5.582 \pm   0.104$ & $  2.696 \pm   0.008$ & $199.5$ & $  8.0$\\
D & s& $5461 \pm  15$ & $  3.690 \pm   0.001$ & $ -0.370 \pm   0.014$ & $  5.875 \pm   0.077$ & $  2.711 \pm   0.006$ & $443.8$ & $ 17.8$\\
 & p& $5445 \pm  21$ & $  3.688 \pm   0.001$ & $ -0.366 \pm   0.012$ & $  5.708 \pm   0.094$ & $  2.688 \pm   0.007$ & $284.9$ & $ 11.4$\\
 & k& $5470 \pm  13$ & $  3.695 \pm   0.000$ & $ -0.384 \pm   0.007$ & $  6.080 \pm   0.063$ & $  2.749 \pm   0.003$ & $1718.1$ & $ 66.1$\\
\\
 & o & $5115 \pm   60$ & $  3.670 \pm   0.008$ & $  0.410 \pm   0.060$ & $  4.646 \pm   0.385$ & $  2.747 \pm   0.082$ & & \\
 & c& $5056 \pm  40$ & $  3.689 \pm   0.001$ & $  0.124 \pm   0.014$ & $  4.927 \pm   0.157$ & $  2.896 \pm   0.007$ & $223.0$ & $  8.9$\\
 & b& $4989 \pm  62$ & $  3.686 \pm   0.001$ & $  0.168 \pm   0.014$ & $  4.595 \pm   0.227$ & $  2.873 \pm   0.005$ & $153.8$ & $  6.4$\\
E & s& $4995 \pm  29$ & $  3.686 \pm   0.001$ & $  0.169 \pm   0.011$ & $  4.622 \pm   0.109$ & $  2.874 \pm   0.006$ & $160.8$ & $  6.7$\\
 & p& $5038 \pm  44$ & $  3.683 \pm   0.002$ & $  0.150 \pm   0.008$ & $  4.692 \pm   0.143$ & $  2.846 \pm   0.009$ & $339.9$ & $ 14.2$\\
 & k& $4845 \pm   9$ & $  3.700 \pm   0.000$ & $  0.174 \pm   0.005$ & $  4.434 \pm   0.035$ & $  2.993 \pm   0.002$ & $6195.3$ & $247.8$\\
\\
 & o & $5120 \pm   55$ & $  3.580 \pm   0.009$ & $ -0.400 \pm   0.080$ & $  4.898 \pm   0.464$ & $  2.814 \pm   0.111$ & & \\
 & c& $5183 \pm  21$ & $  3.575 \pm   0.002$ & $ -0.424 \pm   0.022$ & $  4.920 \pm   0.081$ & $  2.754 \pm   0.011$ & $ 64.8$ & $  3.2$\\
 & b& $5182 \pm  19$ & $  3.574 \pm   0.001$ & $ -0.412 \pm   0.018$ & $  4.896 \pm   0.085$ & $  2.749 \pm   0.011$ & $ 54.9$ & $  2.9$\\
F & s& $5169 \pm  19$ & $  3.578 \pm   0.002$ & $ -0.438 \pm   0.015$ & $  4.943 \pm   0.112$ & $  2.776 \pm   0.014$ & $131.4$ & $  6.9$\\
 & p& $5267 \pm  15$ & $  3.578 \pm   0.002$ & $ -0.487 \pm   0.012$ & $  5.300 \pm   0.067$ & $  2.768 \pm   0.012$ & $188.1$ & $  9.9$\\
 & k& $5561 \pm   8$ & $  3.600 \pm   0.000$ & $ -0.488 \pm   0.005$ & $  7.468 \pm   0.049$ & $  2.948 \pm   0.003$ & $2378.0$ & $118.9$\\
\\
\bottomrule
\end{tabular}

\label{t:obs}
\end{table*}

\subsection{Results and discussion}

Our results are given in Tables~\ref{t:mdl} and \ref{t:obs} for each
star from A to F and each surface correction, labelled by ``c'',
``b'', ``s'', ``p'', ``k'' for the cubic correction, combined
correction, modified Lorentzian, free power law and solar-calibrated
power law, respectively.  Table~\ref{t:mdl} lists the stellar model
parameters, including the relevant free parameters for each of the
surface corrections.  Table~\ref{t:obs} lists the derived parameters,
which can be compared to the row of observational values labelled by
``o''.  The effective temperature $\Teff$ and surface metallicity
$\FeH\st{s}$ are given by \citet{deheuvels2014}; the other parameters
are derived from scaling relations.  Table~\ref{t:obs} also gives the
total misfit $\chi^2$ (i.e. eq.~(\ref{e:chi2})) and reduced misfit
$\chi^2_r=\chi^2/N\st{dof}$, where $N\st{dof}$ is the number of
observations less the number of free parameters.  Figs~\ref{f:cube} to
\ref{f:kbcd} show the frequency differences between the best fitting
models for each star and each surface effect, both before and after
the correction is applied.  Each figure shows the fits for one surface
correction, in the same order as in Tables~\ref{t:mdl} and
\ref{t:obs}.

In all the stars, the inclusion of the mode inertia in the surface
correction is clearly important.  In Figs~\ref{f:cube} to
\ref{f:plaw}, the greatest surface correction is clear for the
low-inertia radial and p-dominated mixed modes.  The
  corrections for the less p-dominated mixed modes fall between zero
  and the trend followed by the radial and p-dominated mixed modes.
These modes have larger inertia and therefore smaller
frequency corrections, and this correctly brings the corrected
frequencies in line with the low-inertia modes.

The precise ages (especially for stars B and D) may come as a surprise
but are easily understood.  In short, the frequencies of the mixed
modes evolve quickly and are measured precisely.  For example, in our
best fit for star A, the mixed mode at $836.040\uHz$ varies
with age at about $1.45\uHz/\Myr$.  Given the observed uncertainty of
$0.015\uHz$, the modelled mode frequency changes by $1\sigma$ in a
little over $10\,000\yr$.  In other words, along a single evolutionary
track, the uncertainty in the age is about $10\,000\yr$.  Most of the
reported uncertainty in age is a result of correlations with other
parameters.

The best-fit models generally reproduce the spectroscopic properties
within about the $2\sigma$ limits of the observations.  Star E is an
exception, almost certainly because of poor modelling (see
Sec.~\ref{ss:settling}).  The models of stars B and D are somewhat hotter
and more metal poor than the spectroscopic determinations suggest,
with the discrepancy nearing the $3\sigma$ level.

\subsection{Gravitational settling without competition}
\label{ss:settling}

Star E appears to have a mass roughly between $1.4$ and $1.6\Msun$.
In such stars, gravitational settling significantly (and sometimes
completely) depletes the stellar surface of its helium and metals,
which is obviously inconsistent with observations.  Once off the
main-sequence, however, the inward-penetrating convection zone partly
restores the initial surface mixture, which allows us to find stellar
models that still have reasonable surface metallicities $\FeH\st{s}$,
though usually still inconsistent with the observed value of
$0.41\pm0.06\dex$.

In reality, our expectation is that at this mass, some other
unmodelled process counteracts the depletion of metals from the
stellar atmosphere.  Possible competing processes have been studied
more extensively in hotter stars---mainly A- and early F-type---and
include radiative levitation \citep[e.g.][]{turcotte1998}, rotation
\citep[e.g.][]{charbonneau1991} or small amounts of mass-loss
\citep[e.g.][]{michaud2011}.  Our models of star E, and thus its
parameters, exclude these competing processes and should be considered
with this in mind.  We have included them partly for completeness and
partly because we still expect that the characteristics most directly
probed by the mode frequencies (e.g. the mean density and surface
gravity) are reasonably accurate, even if the parameters that are
interpreted through stellar models (e.g. the age) are not.

Star D is the next most massive star and potentially also suffers from
the inclusion of overly-efficient gravitational settling.  The initial
helium abundance is slightly lower than typical values from Big Bang
nucleosynthesis of around $0.247$ \citep[e.g][]{cyburt2016} and the
observed surface metallicity is discrepant, though not as severely in
star E.  Star A is also about as massive and the best-fit helium
abundance seems low for such a metal-rich star but the surface
metallicity is poorly constrained by the spectroscopic observations.

\subsection{Which correction is best?}

Though no surface effect correction is obviously superior to the
others, the solar-calibrated power law by \citet{kbcd2008} fits
consistently worse than the other corrections.  In fact, in three
cases (stars A, E and F) the best-fitting model is one with no surface
correction at all.  This result, however, is neither surprising nor an
indictment of the calibrated power-law.  It is calibrated to the Sun,
so while it might be expected to work for Sun-like stars, there is no
reason to expect that it would continue to work for these evolved
stars.  The reasonable results for the uncalibrated power-law show
that a simple power law is not a bad idea, it is only using a solar
calibration out of context that leads to poor results.

In addition, the formulation of \citet{kbcd2008} first rescales the
model frequencies by a factor (originally and here denoted $r$) that
represents the square root of the ratio of the mean densities of the
stellar model and the observed star.  The best-fitting models using
the other corrections imply that the ratio of mean densities is a few
thousandths smaller than one.  Often, the rescaling is what causes the
regression to find that the best-fitting correction is no correction.
No other surface correction uses this factor $r$.

The poor performance of the calibrated power law leads to small
uncertainties by our method of estimation, which is based on surfaces
of constant $\chi^2$ and therefore assumes that the best-fitting model
fits the data reasonably well.  When this is not the case, small
changes in the model parameters increase $\chi^2$ from a large value
to a much larger value and the uncertainty is ultimately
underestimated.  For this reason, the estimated uncertainties of the
surface gravity $\log g$ are sometimes smaller than $0.0005$.  Since
we regard these estimates as unreliable anyway, we have allowed them
to be rounded to zero rather than encumber that column of
Table~\ref{t:obs} with a further significant digit.

Of the other surface corrections, the modified Lorentzian proposed by
\citet{sonoi2015} is the next worst performer in three stars.  In many
cases, much of the poor performance is contributed by the
high-frequency modes, where the modified Lorentzian function fails to
capture the continued increase in the scale of the surface effect.
For example, in stars A and B, the three highest-frequency modes
together contribute $98.3$ and $61.7$ to $\chi^2$, respectively.
Similarly, in Star F, the two discrepant modes between $510$ and
$540\uHz$ together contribute $71.3$ to $\chi^2$.  Over the remaining
modes, the modified Lorentzian surface term performs about as well as
the other corrections.  Put differently, the marginally-worse overall
performance mostly reflects that the modified Lorentzian does not
describe the high-frequency end of the correction very well.  The same
conclusion can be drawn by comparing the modified Lorentzian with the
frequency differences between a solar-calibrated model and low-degree
solar mode frequencies.

The free power law is the next best performer but varies between being
about as good as the cubic and combined terms (stars A and B) and much
worse than the modified Lorentzian (star E).
If the index of the power law were fixed at $3$, the power law
  correction would be similar to the cubic term, except for the
  difference in the treatment of the mode inertia.  In the power law,
  however, the factor $Q$ only corrects for the difference between the
  inertia of the non-radial modes relative to the radial, whereas the
  cubic and combined corrections use the mode inertia $\mathcal{I}$
  without modification.  This difference would be negligible if the
  mode inertia was roughly a power-law in frequency but it is
  generally not, except perhaps over small ranges of frequency.

In terms of the $\chi^2_r$, the combined term usually fits best but it
is difficult to make robust conclusions when comparing models that all
fit quite badly in absolute terms.  Only in star C are the fits
sufficiently good that the cubic and combined terms are significantly
superior.  Our only strong conclusions are that it is inappropriate to
use a solar-calibrated power law to correct for surface effects in
such evolved stars, and that the modified Lorentzian correction does
not match the highest-frequency behaviour of observed surface effects.

\subsection{Uncertainty from choice of correction}

We investigated the level of uncertainty induced by the choice of
surface correction for each star, neglecting the
consistently-discrepant solar-calibrated power law, first by combining
the uncertainties derived for each surface correction in quadrature
and comparing the spread of $1\sigma$ confidence intervals.  In this
sense, the models broadly agree, with the individual $1\sigma$
intervals usually overlapping the combined $1\sigma$ region or
fractionally separated but we note a few exceptions.  First, the
mixing-length parameter for the free power-law correction tends to be
larger than for the other corrections.  In addition, the results for
stars C and E appear to be the least consistent.  We have already
noted the flaws in our models of star E but it is not clear what
disrupts the fits in star C.

We also compared the standard deviations of the central best-fit
values with the individual uncertainties.  There is great variation in
this ratio but we can roughly say that the uncertainty in each
parameter introduced by the choice of surface effect is between one
and two times the uncertainty in each individual fit.  Put
differently, one can say that the choice of surface correction biases
the results relative to one another but usually by no more than the
$2\sigma$ uncertainties.  The cubic and combined terms by
\citet{ball2014} obviously agree better---almost always within the
mutual $1\sigma$ limits---because the cubic correction is a special
case of the combined correction.

As a third comparison of the uncertainties, we combined the results
for each star for each variable in Tables~\ref{t:mdl} and
~\ref{t:obs}, assuming that the results are from uncorrelated normal
distributions.  The overall uncertainties, which now span the full
ranges covered by the fours surface corrections, are usually between
$2$ and $3$ times the uncertainties for the individual fits.  Relative
to the best-fitting values, for the masses $M$, radii $R$ and ages
$t$, we find that the total uncertainties are always less than about
$2$, $1$ and $6$ per cent, respectively.

\begin{table}
\centering
\caption{Mixing-length parameters predicted by linear interpolation in
  the grid of values calibrated to \textsc{Stagger} simulations by \citet{magic2015}.
  The second column gives the mixing-length parameter relative to the value
  of their solar simulation.  The third column gives this ratio multiplied 
  by the solar-calibrated mixing-length parameter for our stellar models,
  $\alpha_{\odot,\mathrm{MESA}}$.}
\begin{tabular}{ccc}
\toprule
Star & $\alpha/\alpha_{\odot,\textsc{Stagger}}$ &  $\times\,\alpha_{\odot,\mathrm{MESA}}$ \\
\midrule
A & $0.982\pm0.033$ & $1.755\pm0.059$ \\
B & $0.961\pm0.005$ & $1.718\pm0.008$ \\
C & $0.944\pm0.105$ & $1.688\pm0.188$ \\
D & $0.978\pm0.014$ & $1.749\pm0.025$ \\
E & $0.933\pm0.061$ & $1.668\pm0.109$ \\
F & $0.880\pm0.064$ & $1.574\pm0.115$ \\
\bottomrule
\end{tabular}
\label{t:stagger}
\end{table}

\subsection{Mixing-length parameters}

Since the landmark work by \citet{ludwig1999}, two- and
three-dimensional radiation hydrodynamics simulations of near-surface
convection have allowed for calibration of the mixing-length parameter
$\alpha$ for cool stars of various types.  In recent years, several
suites of simulations have been used to perform such calibration
\citep[e.g.][]{trampedach2014b, magic2015} and consistently conclude
that, roughly speaking, the mixing-length parameter should decrease as
the effective temperature increases or the surface gravity decreases.

The tables produced by \citet{magic2015} include calibrations for
different metallicities $\FeH$ and allow us to compare our
mixing-length parameters with those predicted from their calibration.
For each star, we generated $10^5$ realizations of $\Delta\nu$,
$\nu\st{max}$, $\Teff$ and $\FeH$ and used them to interpolate in the
data of \citet{magic2015}, using the mixing-length parameters
calibrated to the entropy at the bottom boundary of the simulation.
(Using the calibration to the entropy jump gives nearly identical
results.)  Table~\ref{t:stagger} shows the means and standard
deviations of the samples for each star, divided by the solar value,
which itself depends on the choice of atmospheric model and detailed
implementation of mixing-length theory.  The table also gives the
ratio multiplied by our solar-calibrated value $\alpha_{\odot,\mathrm{MESA}}=1.788$.  The
simulation calibrations consistently suggest that the mixing-length
parameter $\alpha\st{MLT}$ should be less than the solar-calibrated
value.  In our fits, we have found the opposite: $\alpha\st{MLT}$ is
larger than the solar value, mostly falling between $1.8$ and $2.1$.

What causes this discrepancy?  Based on their ability to
reproduce observable features of solar granulation, the simulations
are generally regarded as reasonably realistic, at least for the Sun
\citep[see e.g.][for a review]{nordlund2009}.  In addition, there is
already some observational support for small mixing-length parameters
on the red giant branch.  \citet{piau2011} modelled a range of stars
across the red giant branch ($3.8>\log g>1.5$) for which linear
diameters could be determined to better than 10 per cent, and found
that they were much better able to model the stars with a sub-solar
value for the mixing-length parameter.

Given the strong correlations between the parameters, our
mixing-length parameters are probably being used to adjust the stellar
radii so that the mode frequencies match the observations.  The
incorrect part of the stellar model could then instead be, for
example, the initial helium abundance $Y_0$, the gravitational
settling or another component of the input physics.  It remains to be
seen if this effect persists in asteroseismic models of subgiants and
low-luminosity red giants or if it is specific to the fits presented
here.

\section{Conclusion}
\label{s:conclusion}

We have modelled the individual mode frequencies of six subgiants and
low-luminosity red giants using five different parametrizations of the
surface effects.  The solar-calibrated power-law correction proposed
by \citet{kbcd2008} is clearly unsuitable, consistently producing much
poorer fits and often converging on models without any
surface effect.  This poor performance is not surprising: there is no
reason to expect a solar-calibrated correction to work on evolved
stars that are unlike the present Sun.

The remaining four surface corrections provide fits of similar
quality, with the the combined correction by \citet{ball2014} being
marginally superior for most of the stars.  The best-fitting
parameters of these four sets of fits are generally mutually agreeable
within the $2\sigma$ uncertainties, with the exception that the
parameters found with the modified Lorentzian or free power law
occasionally disagree at slightly more than the mutual $3\sigma$
level.  Put differently, the systematic uncertainty is roughly twice
the statistical uncertainty in our fits (though it varies for
different parameters).  For the masses, radii and ages of the stars,
the total fractional uncertainties are always smaller than about $2$,
$1$ and $6$ per cent, respectively.

Finally, we note that the present results demonstrate that the
individual mode frequencies of these subgiants and low-luminosity red
giants can be used to constrain models, in much the same
fashion as is now commonplace for main-sequence dwarfs.  Our modelling
procedure is only marginally different from what has already been used
for dwarf solar-like oscillators, almost entirely in allowing shorter
timesteps to resolve the rapid evolution of the mixed mode
frequencies.  Even given the additional uncertainty introduced by the
surface correction, the model parameters of these stars can help to
precisely constrain the physics of stellar interiors.

\begin{acknowledgements}
  The authors acknowledge research funding by Deutsche
  Forschungsgemeinschaft (DFG) under grant SFB 963/1 ``Astrophysical
  flow instabilities and turbulence'', Project A18.  LG acknowledges
  support by the Center for Space Science at the NYU Abu Dhabi
  Institute under grant G1502.
\end{acknowledgements}


\end{document}